\documentstyle[12pt,epsfig]{article}
\begin{document}

\begin{titlepage}

\title{\small\raggedleft{astro-ph/9702038}\\
\small\raggedleft{DAMTP-R/9707}\\
\LARGE\center\bf High-Resolution Analysis of Cold and Hot Dark Matter in
Cosmic String Wakes} 
\author{A. Sornborger,\\
        DAMTP,\\
        Cambridge University,\\
        3 Silver Street,\\
        Cambridge, CB3 9EW,\\
        UK}
\date{February 3, 1997}
\end{titlepage}

\maketitle

\begin{abstract}
We study the accretion of cold and hot dark matter onto a cosmic
string wake using a high-resolution numerical simulation. We verify
previous analytical calculations predicting the radius of bound matter
around wakes and inflow velocities of the dark matter, as well as
assumptions about the self-similarity of the solution. In cold wakes,
we show that self-similarity is approached quickly after a `binding'
transition. Hot wakes approach self-similarity rapidly once the free
streaming `pressure' falls below a critical value and accretion
begins. We also analytically calculate the size of the overdensity in
wakes with cold dark matter and compare the results to simulations. We
remark that the results derived may be used in testing gravitational
numerical codes in the non-linear regime.
\end{abstract}

\section{Introduction}
Large-scale structure formation in the cosmic string model is not yet
well understood, particularly in the non-linear regime. Until
recently, only linear analytical treatments have been available for
approximating wake characteristics. Rees $^{\cite{analbaryo}}$ has
analytically calculated some of the non-linear effects of baryons in
wakes, particularly at the time of formation.

There have also been some recent numerical attempts at understanding
the matter distribution better. Shellard and Avelino
$^{\cite{linwake}}$ have developed Zel'dovich techniques including the
source terms from cosmic strings, but this work is only valid on large
scales where the perturbations are still linear. Sornborger,
et. al. $^{\cite{baryowake}}$ have simulated non-linear evolution of
baryons and dark matter in cosmic string wakes, but their simulations
had small grids, and therefore could not resolve non-linear features
in the matter distribution well. Their simulations also have the
drawback that boundary conditions come into play typically by a
redshift of $8$, and therefore assumptions have to be made about the
scaling of the matter distribution (i.e. that it is self-similar).

Therefore, what is needed is a confirmation of both linear and
non-linear results and evolution of the wakes until the present in
order to verify that the assumptions about the growth of the matter
distribution in the wake are correct. We would like more accurate
analytical expressions for the size of wake overdensities to compare
with observations. Also, since cosmic strings with cold dark matter
generate too much small scale structure $^{\cite{coldspect}}$, an
investigation of the effects of the accretion of hot dark matter would
be desirable.

In this paper, we have done high-resolution one-dimensional numerical
simulations yielding precise comparisons with the Zel'dovich
approximation in the linear regime. In addition, we have extended the
Zel'dovich approximation, in a slightly modified guise, to make
predictions for the width of the overdensity in the wake into the
non-linear regime. We have also extended our simulation to include the
effects of hot dark matter.

The cosmic string model of large-scale structure is motivated by the
fact that supermassive strings are a common product of a GUT scale
phase transition. A string network is formed at the phase
transition. The energy carried by the defects curves spacetime, and
thus seeds structure formation. Simple arguments
$^{\cite{stringscale}}$, verified by numerical simulations, show that
the string network rapidly evolves to a scaling solution. That is, the
number of long strings per horizon volume is statistically constant
(i.e. when we average over many volumes).

Numerical simulations have shown that cosmic string loops have a
typical radius at the limit of resolution of cosmic string network
simulations $R \sim 10^{-2}H^{-1}$, whereas there are $\sim 10 - 30$
horizon sized long strings per particle horizon volume in the matter 
era $^{\cite{stringsim1}, \cite{stringsim2}, \cite{stringsim3}}$. This
indicates that the primary structure formation mechanism in the cosmic
string model is in the form of sheetlike wakes formed in the path of
long strings.

The matter perturbation caused by a moving long cosmic string comes
from two conceptually distinct components of the gravitational field
$^{\cite{stringgrav1}, \cite{stringgrav2}, \cite{stringgrav3}}$: a
velocity kick imparted to particles toward the plane traversed by the
cosmic string due to converging geodesics in the conical spacetime
around the string, and a Newtonian field due to the effective mass of
small-scale structures such as waves and kinks moving on the
string. In linear theory, the comoving velocity impulse toward the
wake imparted to a particle after a string has passed by is
$$v_{kick} = -\frac{2 \pi G (\mu_{eff} - T)}{v_s \gamma_s} - 4 \pi G
\mu_{eff} v_s \gamma_s$$
where $G$ is Newton's constant, $\mu_{eff}$ is the effective
mass-per-length of the string, $v_s$ is the velocity of the string,
$\gamma_s$ is the relativistic gamma factor $\gamma_s = (1 -
v_s^2)^{-1/2}$ and $T$ is the tension on the string which obeys an
effective equation of state $\mu_{eff} T = \mu_0$. For strings formed
at a GUT phase transition, $G\mu_0 \sim 10^{-6}$. In the expression
for $v_{kick}$, the first term comes from the Newtonian field and the
second from the deficit angle of the spacetime. Thus, we can see that
for fast moving strings ($v_s \sim 1$) the impulse from the Newtonian
field is negligible compared to that from the deficit angle in the
conical spacetime around the string.

To study matter accretion in cosmic string wakes, planar symmetry is
often assumed as a first order approximation for ease of
calculation. That is, we assume the wake to be formed by a cosmic
string in the regime where the string velocity is relativistic (a good
approximation) and the string straight (a good approximation if we
neglect small scale structure on the strings, which is at present not
well understood $^{\cite{stringsim1},
\cite{stringsim2}, \cite{stringsim3}}$, but can be hoped to be of the
order of the cosmic string loop scale $< 10^{-2} H$). Thus, we
neglect the Newtonian field of the string, and we assume that the
string has passed through the volume of interest without deviating
from being straight, such that the inflow is effectively
one-dimensional. In this scenario, two matter streams are formed about
the axis where the cosmic string has passed, each with identical
velocities toward the axis.

Analytical calculations using the Zel'dovich approximation
$^{\cite{zel}, \cite{cosstrwake}}$ and assuming planar symmetry have
shown that the wake width should grow as $t^{2/3}$ in the matter
dominated era. In these calculations, the comoving wake width was
defined as the physical radius $h$ from the wake at which the particle
turns around ($\dot h = 0$). For hot dark matter, the width of the 
wake has been calculated including the effects of the larger phase
space volume occupied by dark matter. These calculations predict that
hot dark matter wakes laid down at a $t_{eq}$ will be of a size the
same order as that of cold dark matter wakes laid down at the same
time.

Since the perturbation overdensity $\delta \rho/ \rho = 1$ in the
wake immediately following passage of the string, the solution is
non-linear \it ab initio \rm inside the overdensity. For times $t \gg
t_i$, where $t_i$ is the formation time of the wake, it can be shown
that matter accretion has no preferred length or time scales, thus the
matter distribution in the wake approaches self-similarity. 

In this paper, we study planar symmetric wakes numerically and derive
and compare where we can with analytical results. We have developed a
high-resolution one-dimensional particle-mesh code to study hot and
cold dark matter accretion in cosmic string wakes in the non-linear
regime. The calculations are done assuming a matter dominated
background.

In the next section, we analytically calculate the size of the
overdensity in the wake. Then, in section 3, we present details of
the initial and boundary conditions and methods used for the
simulations. In section 4, we give results of the numerical simulation
and compare these with our own and other previous analytical
results. Finally, we discuss our results and present our conclusions.

\section{The Width of the Wake Overdensity}
The equations in Lagrangian coordinates for a dark matter particle
interacting with a gravitational field in an expanding universe in
one-dimension are (see for instance \cite{pmeq}):
\begin{equation}
\dot v + 2\frac{\dot a}{a} v = - \frac{1}{a^3} \partial \phi
\label{moment} 
\end{equation}
\begin{equation}
\dot r = v \label{pos}
\end{equation}
\begin{equation}
\partial^2 \phi = 4 \pi G (\rho - \bar\rho). \label{grav}
\end{equation}
We are working in comoving coordinates where $v = (1/a) v_p - (\dot
a/a) x_p$ is the comoving particle velocity (subscript $p$ indicates
physical quantities), $a = a(t)$ is the scale factor, $\phi = a\phi_p
+ (a^2/2) \ddot a x_p^2$ is the comoving gravitational potential,
$r = r_p/a$ is the comoving particle position, $\rho = a^3
\rho_p$ is the comoving mass density, and $\bar\rho$ is the average
comoving background density. An overdot indicates derivative with
respect to time, $\partial$ denotes a spatial derivative and we
normalize the scale factor to unity at the present time.

Using the Zel'dovich approximation, it has been calculated that the
comoving infall velocity of particles outside the wake is (see, for
instance $^{\cite{cosstrwake}}$)
\begin{equation}
v_{stream} = \frac{2}{5} v_i \bigg (\frac{t}{t_i}\bigg )^{-\frac{1}{3}} 
     + \frac{3}{5} v_i \bigg (\frac{t}{t_i}\bigg )^{-2} \label{streamvel}
\end{equation}
where subscript $i$ indicates the initial value of the variable
(i.e. wake formation).

As mentioned in the introduction, analytical calculations using the
Zel'- dovich approximation have estimated the width of the wake by
calculating the radius at which the {\it physical} velocity of a
particle goes to zero with respect to the plane of the wake
$^{\cite{zel}}$. However, the overdensity is bounded by the radius at
which particles turn around in comoving coordinates, thus forming the
first caustic in the density distribution. We refer to this radius as
the `second turnaround radius'. We can calculate this radius as
follows:

We study a particle lying initially at the symmetry axis just as it is
kicked by the passage of a string. We follow the particle's trajectory
after the kick occurs. A particle at the axis when the kick occurs
will travel away from the symmetry axis and be at the front edge of
the overdensity. Thus the entire overdensity is behind it and the
density can be given as a time dependent part times a delta-function
distribution:
\begin{equation}
\rho = \bar\rho + \rho(t) \delta(x)
\end{equation}
We can then reduce equations \ref{moment}, \ref{pos} and \ref{grav}
to
\begin{equation}
\dot{(a^2 v)} = \mp\frac{1}{a} 2 \pi G \rho(t) \label{moment2}
\end{equation}
\begin{equation}
\phi = 2 \pi G \rho(t) |x| \label{grav2}
\end{equation}
\begin{equation}
\dot r = v \label{pos2}
\end{equation}
where $-$ indicates a particle moving away from the wake and $+$ is a
particle moving towards the wake.

We know the rate at which the particles are falling into the wake from
equation \ref{streamvel}. Thus, the overdensity is growing as $\rho(t)
= 2 \rho_b x_{stream}(t)$, where $x_{stream}$ is the distance
particles have fallen towards the wake, given by the stream
velocity integrated from the initial time to time $t$. That is
\begin{equation}
\rho(t) = \frac{6}{5} v_i \rho_b t_i \bigg [
\bigg ( \frac{t}{t_i} \bigg )^\frac{2}{3} - \bigg ( \frac{t}{t_i}
\bigg )^{-1} \bigg ]
\end{equation}

Using eq. \ref{moment2}, and $\rho_b = 1/6 \pi G t_0^2$ and $v_{kick_p}
\equiv v_{kick} = v_i (t_i/t_0)^{2/3}$, where $t_0$ is the
present time, and $t_i$ is the time of wake formation, we find the
comoving particle velocity
\begin{equation}
v(t) = v_{kick} \bigg (\frac{t_0}{t_i}\bigg)^\frac{2}{3} \bigg(
\frac{t}{t_i} \bigg )^{-\frac{4}{3}} \Bigg [1 \mp \bigg(\frac{2}{5}
\bigg (\frac{t}{t_i} \bigg ) + \frac{3}{5} \bigg ( \frac{t}{t_i} \bigg
)^{-\frac{2}{3}} - 1 \bigg ) \Bigg ] \label{partvel}
\end{equation}
which is consistent with that found from the Zel'dovich approximation;
and the particle trajectory
\begin{equation}
x(t) = v_{kick} \bigg(\frac{t_0}{t_i}\bigg)^\frac{2}{3} t_i \Bigg[ -3
\bigg (\frac{t}{t_i}^{-\frac{1}{3}} - 1\bigg ) \mp
\bigg[ \frac{3}{5} \frac{t}{t_i}^\frac{2}{3} - \frac{3}{5}
\frac{t}{t_i}^{-1} + 3 \bigg (\frac{t}{t_i}^{-\frac{1}{3}} - 1 \bigg )
\bigg] \Bigg] \label{parttraj}
\end{equation}

These results are exact until the particle's trajectory crosses with
that of another particle.

Second turnaround occurs at $t = t_{turn}$, given by $v(t_{turn}) =
0$. Using eq. \ref{partvel} we find $t_{turn} = b t_i$, where $b =
4.452$. This gives a second turnaround radius $x(t_{turn}) = v_{kick}
(t_0/t_i)^{2/3} t_i \, d$. Where $d = 0.8638$

As we shall see from the numerical results, the solution rapidly
enters a self-similar growth phase, with turnaround radius scaling as
$x_{turn} \sim t^{2/3}$. However, there is an intermediate time when
the wake `binds'. During this time, the second turnaround radius
decreases, then increases once more at the onset of $t^{2/3}$
growth. We can calculate the radius in the growing phase (after
binding) using the approximation outlined above, but now, we consider
a particle falling in from one side of the wake, crossing the symmetry
axis, and flowing out the other side of the wake. This calculation is
an extension of the Zel'dovich approximation into the non-linear
regime since the particle trajectory we consider will cross other
particle trajectories. However, the net effect of the gravitational
field should be the same, as long as the particle turns around outside
of the rest of the particles in the wake. Keeping only growing mode
terms, we find the turnaround radius
\begin{equation}
x_{turn} = \frac{2}{5} f v_{kick}
\bigg(\frac{t_0}{t_i}\bigg)^\frac{2}{3} t_i
\bigg(\frac{t_{turn}}{t_i}\bigg)^\frac{2}{3} \label{xturn}
\end{equation}
Including damping mode terms shifts $f$. Simulation results given
below give a good match to this expression with $f = 6.7$.

\section{Numerical Scenario}
We use standard particle-in-cell (PIC) techniques to simulate dark
matter in an expanding universe. PIC techniques use particles which
statistically represent mass in a fluid, in our case the fluid is
collisionless. The particle masses are interpolated onto a mesh giving
a mass density. Next, the potential is solved for by first Fourier
transforming the density field, then solving an algebraic equation
for the potential in Fourier space, then retransforming back to
configuration space. The discrete derivative of the potential then
gives the gravitational force, which is interpolated back to the
particle positions. Once the force is known on each particle, a
staggered leapfrog algorithm updates the particle locations and
velocities and we continue updating particle positions and velocities
in time. We check that energy is conserved by our code to a fraction
of a percent.

As initial conditions for the simulation, we give an ingoing velocity
to two matter streams on either side of a symmetry axis and we set
periodic boundary conditions. Our simulation has typical grids of
$2048$ zones. For cold dark matter, we use the same number of 
particles as gridzones. However, for hot dark matter, in order to
reduce shot noise from the thermal velocity distribution, we use 512
particles per zone. This reduces spurious statistical fluctuations in
the overdensity to $\delta \rho/\rho \simeq 0.05$.

For hot dark matter, the physical thermal velocity peaks at $t_{eq}$
and then decreases as the scale factor during the matter dominated era
$^{\cite{zel}}$. The mean physical thermal velocity at $t_{eq}$ is
given by the neutrino mass, which is determined by requiring that
neutrinos make up the critical energy density for an $\Omega = 1$
universe. We find
\begin{equation}
v_{eq_p} = T_{\nu, eq}/m_\nu \simeq 0.05
\end{equation}
where $v_{eq_p}$ is the physical thermal velocity at $t_{eq}$,
$T_{\nu,eq}$ is the neutrino `temperature' at $t_{eq}$ and $m_\nu$ is
the neutrino mass. Thus, in hot dark matter simulations, a Maxwell -
Boltzmann distribution of velocities peaked about this thermal
velocity is then added on top of the streaming velocities of the
particles.

Throughout the paper, we assume a critical universe with $\Omega = 1$
and $h = 0.5$.

\section{Simulation Results}
\begin{figure}
\centerline{\psfig{figure=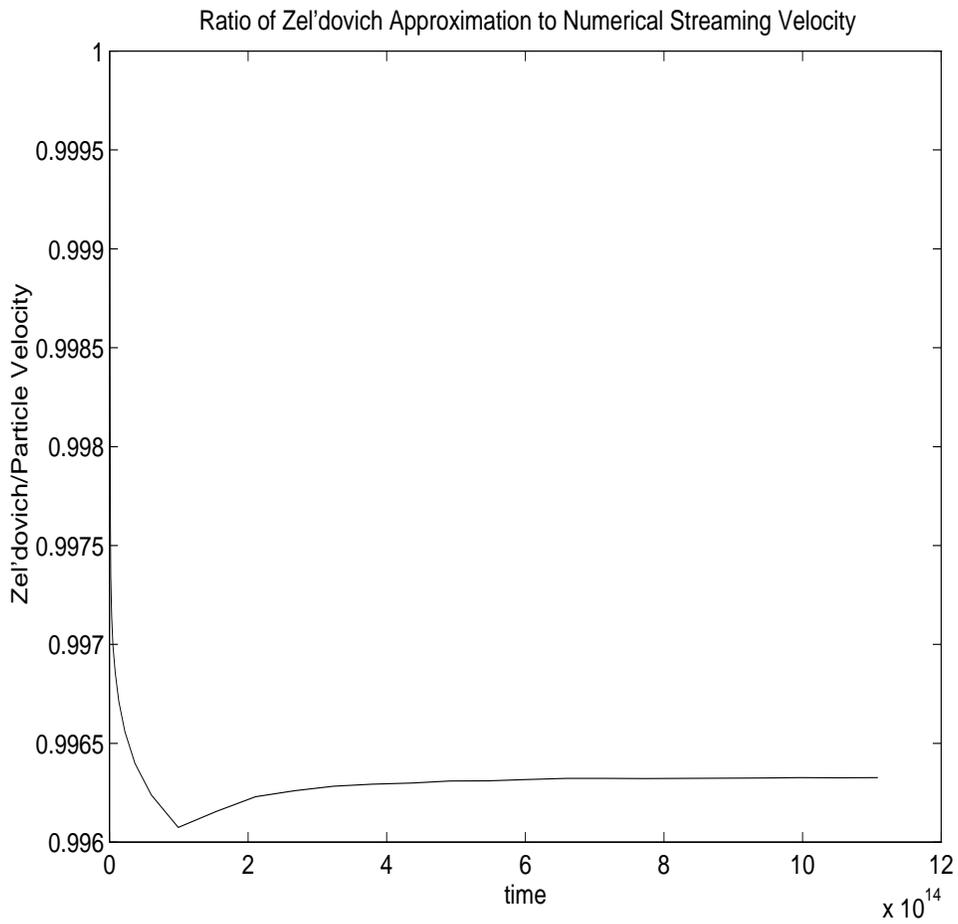,height=5.0in,width=5.0in}}
\caption{A plot of the ratio of the analytical infall velocity to the
numerical result. After an initial correction, the infall velocity is
correct to $\%0.4$.
\label{zeldov}}
\end{figure}

We have checked the infall velocity by comparing with the infall
velocity from the Zel'dovich approximation. A plot of the ratio of the
Zel'dovich approximation velocity with the numerical velocity is given
in figure \ref{zeldov}. Notice that the maximum difference is by a
factor of $0.4 \%$.

\begin{figure}
\centerline{\psfig{figure=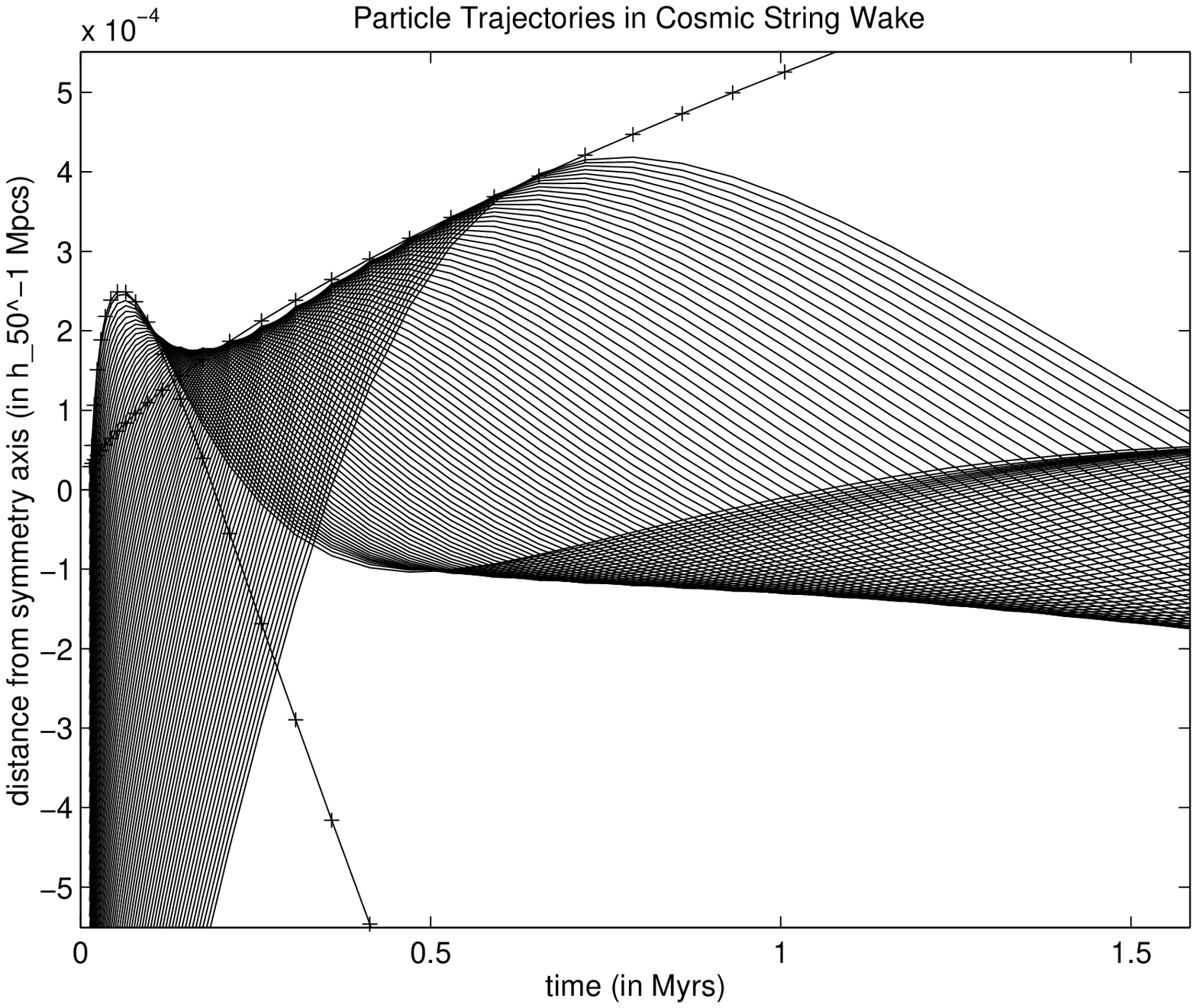,height=5.0in,width=5.0in}}
\caption{A plot of the trajectories of a number of particles
compared with analytical results. In the first stage of growth of the
wake the leading particle's trajectory is enveloped by the analytical
trajectory for a particle being kicked away from the symmetry axis. In
the second stage, where particle turnaround grows as $a(t)$, the
turnaround radius is enveloped by the analytical $t^{2/3}$ power
law. In this figure, solid curves with $+$ signs indicate analytical
results, other solid curves are numerical results.
\label{envelope}}
\end{figure}

\begin{figure}
\centerline{\psfig{figure=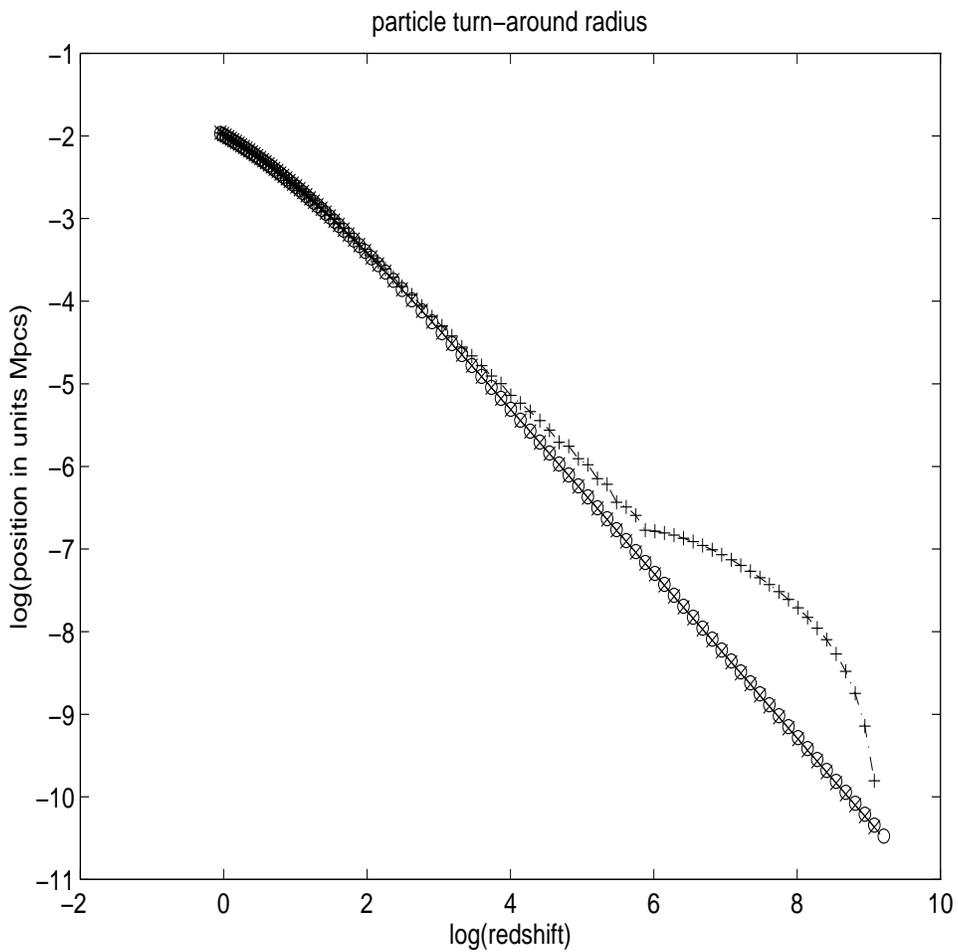,height=5.0in,width=5.0in}}
\caption{Here, particle turnaround (ie. the width of the wake)
(dash-dot line with plus signs) is plotted versus the analytically
calculated growth law (solid line with circles and crosses) well into
the evolution of the wake. Note the close correspondance over the
final $3$ orders of magnitude in redshift.
\label{turnaround}}
\end{figure}

As mentioned above, the analytical result for the `second turnaround'
is exact until just after the point where a particle crosses the path
of another particle. In figure \ref{envelope}, we plot the analytical
result versus particle trajectories. The discrepancy can be seen once
the particle falls back toward the symmetry axis. It is interesting
to note that the wake width does not grow as $t^{2/3}$ until after
`binding' occurs. The solution follows a transient from the formation
stage to the growing stage, resulting in a smaller wake than would be
predicted if one assumed the onset of self-similarity directly after
the first particle falls back toward the symmetry plane. We also plot
the growth ($\sim a(t)$) given in equation \ref{xturn} from the point
where this second stage in the evolution of the wake takes over. In
figure \ref{turnaround} we plot the second turnaround radii
(analytical and numerical) for the evolution of the wake until late
times. The analytical curve matches well with the numerical wake
radius relatively quickly (notice that the plot axes are
logarithmic). If we extrapolate the curve to time $t_0$, we find a
wake size of about $2 x_{turn} = 0.55h_{50}^{-1} Mpc$ for $z_i =
10000$ and $v_s = 0.5$.

\begin{figure}
\centerline{\psfig{figure=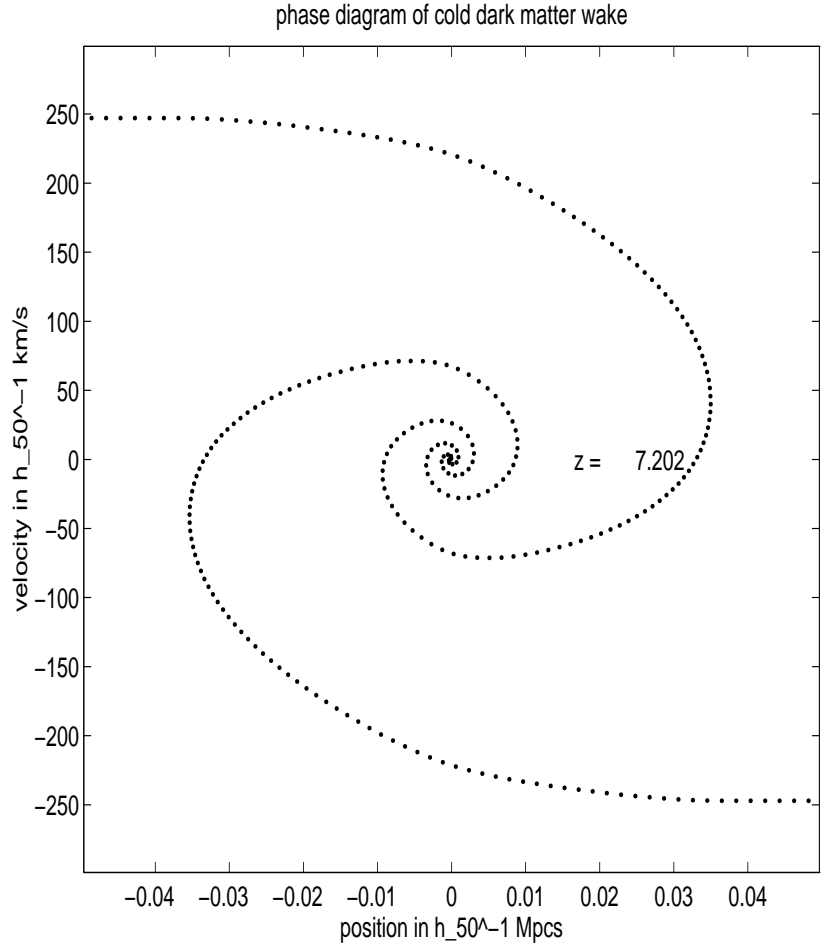,height=5.0in,width=5.0in}}
\caption{A plot at a late time ($z = 7.202$) of the particle phase
space of a cold dark matter wake formed at $z_i = 10000$ with
$v_{kick} = 0.5c$. Each turnaround in the phase space trajectory
corresponds to a peak in the wake overdensity.
\label{coldwake1}}
\end{figure}

\begin{figure}
\centerline{\psfig{figure=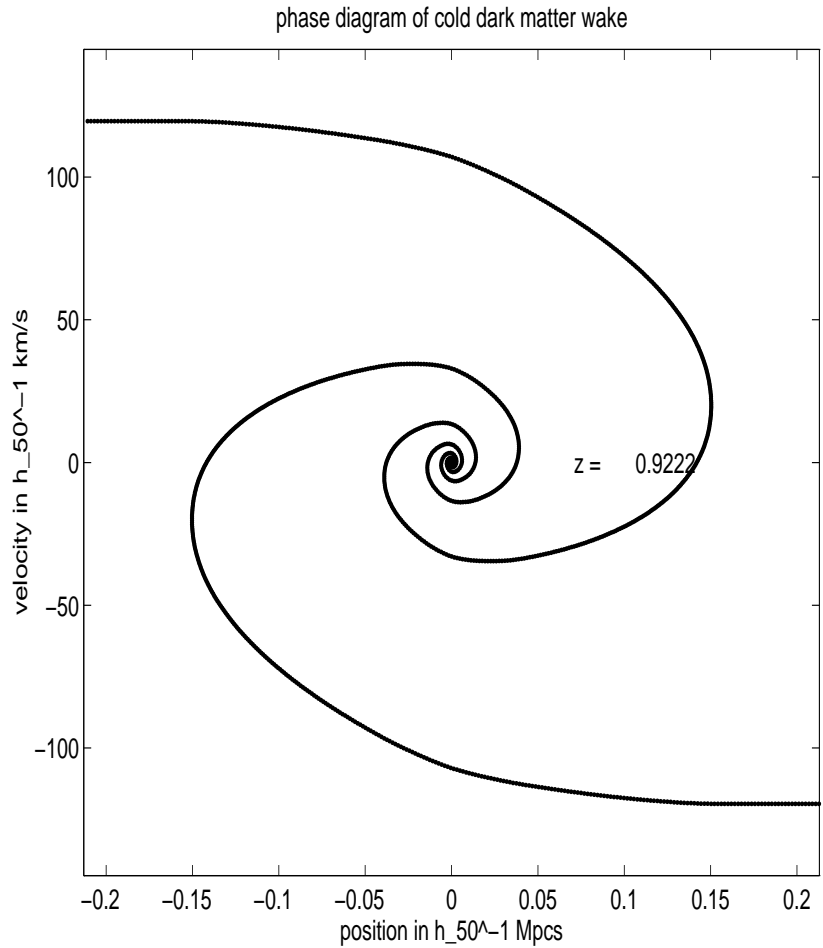,height=5.0in,width=5.0in}}
\caption{A plot at a late time ($z = 0.922$) of the particle phase
space of a cold dark matter wake formed at $z_i = 10000$ with
$v_{kick} = 0.5c$. Notice the self-similarity with the previous
figure. 
\label{coldwake2}} 
\end{figure}

In figures \ref{coldwake1} and \ref{coldwake2}, we plot the phase
space of the particles in a cold dark matter simulation with $z_i =
10000$ and $v_i = 0.5c$. The velocity axis scales as $a^{-1/2}$ and
the position axis scales as $a(t)$ in order to make the
self-similarity evident. The phase space is similar to that shown in
Fillmore and Goldreich ${\cite{selfsim}}$ in their figure 2. The main
difference comes from the velocities outside the non-linear region. In
the phase space of a cosmic string wake, the streaming velocity
outside the non-linear region is constant. This is because the
velocity perturbation from the string is due to the conical deficit
angle which extends approximately to the Hubble radius. Thus, in the
Newtonian approximation, all particles have the same velocity,
regardless of distance from the symmetry axis.

\begin{figure}
\centerline{\psfig{figure=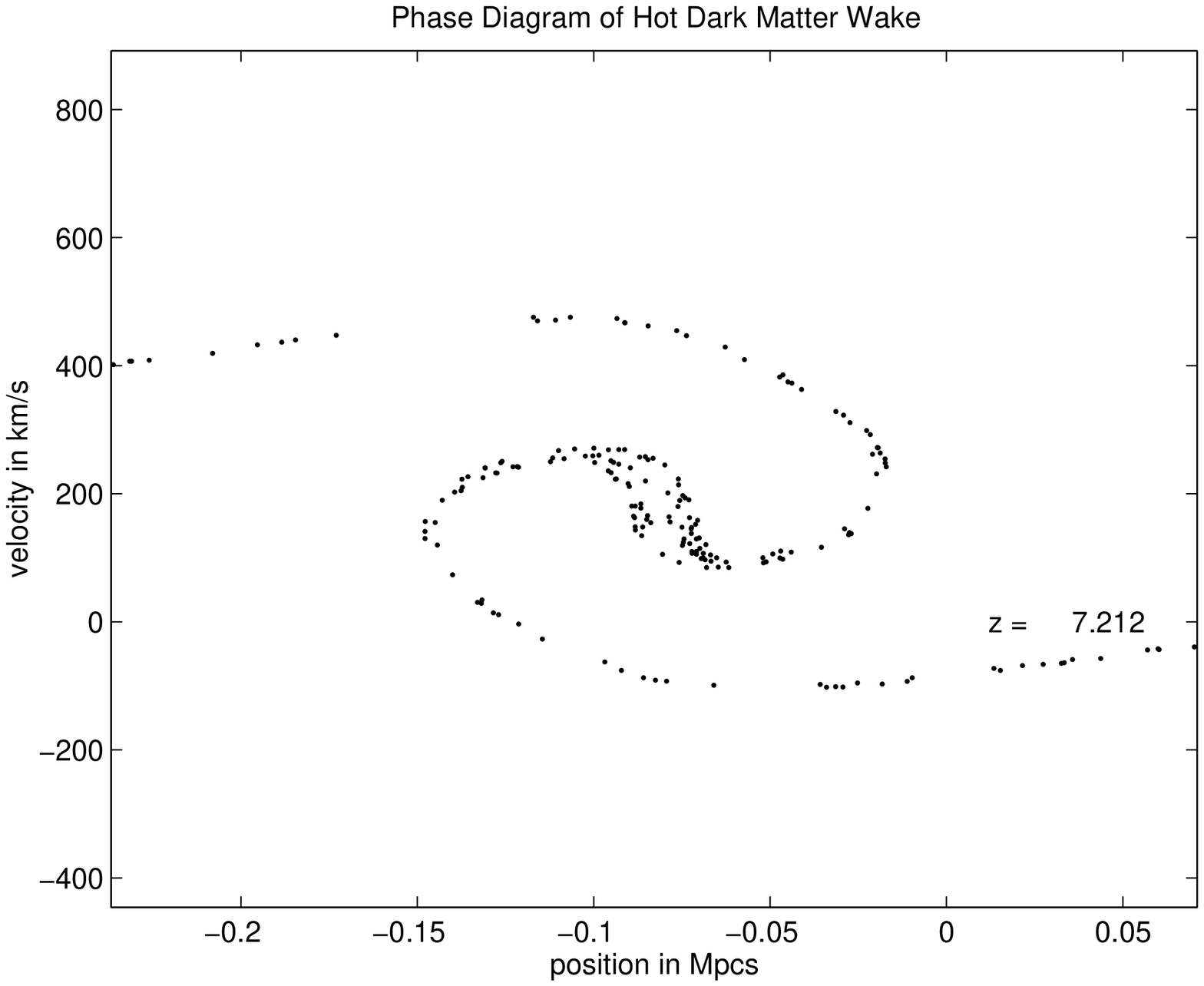,height=5.0in,width=5.0in}}
\caption{A plot at a late time ($z = 7.212$) of the particle phase
space of a hot dark matter wake formed at $z_i = 10000$ with $v_{kick}
= 0.5c$. Each turnaround in the phase space trajectory corresponds to
a peak in the wake overdensity. The position of only one in $512$
particles is plotted.
\label{hotwake}}
\end{figure}

In figure \ref{hotwake}, we plot the phase space of the particles
in a hot dark matter simulation with $z_i = 10000$ and $v_i =
0.5c$. The phase space looks similar (except for a certain raggedness,
which is from shot noise) to that of the cold dark matter wake at late
times. In simulations, we find that accretion does not begin until the
free-streaming length is smaller than the width of the wake. Before
this time, the effective pressure of the neutrino gas keeps structure
from forming on small scales. 

The neutrino free-streaming length grows until $t_{eq}$, then decays
as $a(t)^{-1/2}$. We can approximately calculate the time when wakes
formed at $t_{eq}$ should begin to accrete by finding the time when
the particle turnaround (calculated above for cold dark matter) is
equal to the free-streaming length. This calculation gives a result
\begin{equation}
t_{accrete} \simeq 4.5 \times 10^{24} \frac{1}{b v_{kick}}
\bigg(\frac{b t_{eq}}{t_0}\bigg).
\end{equation}
Thus, accretion should begin at $z_{accrete} \sim 30$.

Numerically, we define a wake to be accreting once the overdensity
$\delta \rho/\rho = 1$. We find hot wakes formed at $t_{eq}$ beginning
to accrete in our numerical simulations at $z_{accrete} \simeq
30$ and second turnaround achieved at $z_{turn} \simeq 8$. After
second turnaround the phase space of hot wakes resembles that of cold
wakes with the overdensity rapidly becoming self-similar and growing
roughly as $a(t)$. We cannot make more quantitative statements due
to the growth of spurious perturbations from statistical fluctuations
in the initial conditions. However, after the onset of accretion in
hot dark matter wakes, the evolution proceeds much as it would for
cold dark matter with a rapid approach to self-similarity.

\section{Discussion and Conclusions}
In this work, we have presented results from a high-resolution
numerical simulation of the evolution of the hot and cold dark matter
distribution in cosmic string wakes. We have also extended the
Zel'dovich approximation and found expressions for the size of the
dark matter overdensity in wakes. We have compared these and previous
results derived using the Zel'dovich approximation to our numerical
results and find good agreement. 

The most massive wakes today are those that have the longest time to
accrete. These wakes are formed at $t_{eq}$. Our simulations give a
wake width of $1.17h_{50}^{-1} Mpc$ for wakes formed at $z_{eq}$. Hot
dark matter wakes are kept from accreting until the mean free path of
the neutrinos falls below the size scale of the wake. The redshift
that this occurs for a wake formed at $t_{eq}$ with $v_{kick} = 0.5c$
is $z_{accrete} \sim 30$. We find good agreement with this prediction
in our hot dark matter simulations, with accretion beginning at the
time calculated from rough arguments.

In a cold dark matter wake, our results show two distinct stages of
behavior of the wake width: the first stage is a brief epoch of
essentially linear growth and contraction immediately following wake
formation, and a second stage in which the overdensity begins to grow
as particles accrete onto the wake. The transition between the linear
stage and the growing stage is an interesting non-linear
phenomenon. The fast approach to non-linearity makes the cosmic string
wake a useful testbed for n-body codes.

To get an understanding of the cosmological implications of these
results, we use data from network simulations. Cosmic string
velocities from simulations are distributed with a broad peak in the
distribution at about $v_s = 0.3$ on horizon scales. Thus, the average
cold string wake formed at time $t_{eq}$ will form a wake of mass $3.2
\times 10^{-8}$ of the mass in the horizon at $t_0$. Combined with
network simulation results (mentioned above) that there are $\sim 10 -
30$ strings per horizon volume, all of the matter in a cold dark
matter universe would be non-linearly accreted at present. Some
caveats to this result are that our simulations assume a matter
dominated universe, and that coarse-graining in network simulations
may lead to an incorrect estimate of the number of strings per horizon
${\cite{netgrain}}$. Nevertheless, a significant fraction $\sim 1/6$
of the matter in the universe would still be non-linear even if there
were only $\sim 1$ string per horizon.

In the analogous hot dark matter universe, significantly less matter
will be non-linear due to the late accretion time for hot wakes. Hot
wakes formed at $t_{eq}$ by cosmic strings with velocities $v \sim
0.5c$ begin to accrete at $z \sim 30$. Early structure formation is
therefore not a problem in the HDM plus cosmic string model (strings
with higher velocities will form wakes that accrete even earlier). If
we use the cold dark matter expressions for wakes forming at this
redshift, we find that a fraction $0.002$ of the matter in the
universe is in hot wakes. This fraction will go up if we integrate
over all wakes. Also, recent work on the string network scaling
solution shows that the length in long strings is roughly two orders
of magnitude higher in the radiation era than in the matter era
$^{\cite{wakescale}}$. This work shows that there is a relatively long
transition between the two scaling regimes, and thus, there will be
more long strings than previously thought around $t_{eq}$. This effect
will again push the amount of accreted matter upwards.

\section{Acknowledgements}
I would like to thank R. Brandenberger for helpful discussions. This
work has been supported by UK PPARC grant GR/K29272.

\end{document}